%% file: ms_revised.tex
\documentclass[iop]{emulateapj}
\usepackage{color}
\usepackage{natbib}
\newcommand{\Teff}{T_{\rm eff}}
\newcommand{\Teq}{T_{\rm eq}}
\newcommand{\Tint}{T_{\rm int}}
\newcommand{\Leff}{L_{\rm eff}}
\newcommand{\Leq}{L_{\rm eq}}
\newcommand{\Eint}{E_{\rm int}}
\newcommand{\ME}{M_{\oplus}}
\newcommand{\RE}{R_{\oplus}}
\newcommand{\MJ}{M_{\rm J}}
\newcommand{\RJ}{R_{\rm J}}
\newcommand{\Zsol}{Z_{\odot}}
\newcommand{\gccm}{g$\,$cm$^{-3}$}

\newcommand{\Tcore}{T_{\rm c}}
\newcommand{\Mcore}{M_{\rm c}}
\newcommand{\Rmean}{R_{\rm m}}
\newcommand{\Ptrans}{P_{\rm 1-2}}
\newcommand{\fig}{Fig.$\:$}

\newcommand{\sect}{\S$\:$}
\newcommand{\nI}{\mbox{$\lambda$}}

\shorttitle{Jupiter models with improved ab initio H EOS}
\shortauthors{Nettelmann, et al.}

\begin{document}

\title{Jupiter models with improved ab initio hydrogen EOS (H-REOS.2)}

\author{N. Nettelmann, A.~Becker, B. Holst\altaffilmark{1}, 
and R. Redmer}
\affil{Institut f\"ur Physik, Universit\"at Rostock, D-18051 Rostock, Germany}

\altaffiltext{1}{CEA, DAM, DIF, F-91297 Arpajon, France}

\slugcomment{Accepted to ApJ, February 24, 2012}

\begin{abstract}
The amount and distribution of heavy elements in Jupiter gives indications on the process of
its formation and evolution. Core mass and metallicity predictions however depend on the 
equations of state used, and on model assumptions.
We present an improved ab initio hydrogen equation of state, H-REOS.2 
and compute the internal structure and thermal evolution of Jupiter within 
the standard three-layer approach. The advance over our previous Jupiter models with
H-REOS.1 by Nettelmann et al.~(2008) is that the new models are also consistent with the observed 
$\gtrsim 2$ times solar heavy element abundances in Jupiter's atmosphere. 
Such models have a rock core mass $\Mcore=0$--$8\:\ME$, total mass of heavy elements $M_Z=28$--$32\:\ME$, 
a deep internal layer boundary at $\geq 4$~Mbar, and a cooling time of 4.4--5.0~Gyrs when assuming 
homogeneous evolution. We also calculate two-layer models in the manner of Militzer et al.~(2008) and find 
a comparable large core of 16-21$\:\ME$, out of which $\sim 11\:\ME$ is helium, 
but a significantly higher envelope metallicity of 4.5$\times$ solar. 
According to our preferred three-layer models, neither the characteristic frequency 
($\nu_0\sim 156\,\mu$Hz) nor the normalized moment of inertia (\nI$\sim 0.276$) are sensitive 
to the core mass but accurate measurements could well help to rule out some classes of models. 
\end{abstract}


\section{Introduction}

Numerous discoveries of unusual hot Jupiters, in particular those that appear to be 
abnormally inflated (HD$\:$209458b; \citealp{Baraffe+03}), extremely metal-rich (HAT-P-2b; 
\citealp{Leconte+09}), tidally disrupted (WASP-12b; \citealp{Li+10}), that are thought to have formed 
by gravitational instability (HR$\:$8799b-e; \citealt{Baruteau+11}), or that even allow 
to study non-equilibrium chemistry in irradiated atmospheres
(HD$\:$189733b; \citealt{Fortney+10}), have drawn some attention away from Jupiter. 

While models for exoplanets often span a variety of possible solutions 
due to the small number of observables characterizing the planet such as radius, mass, age, 
and their rather large error bars, with Jupiter we face the opposite problem that many 
models fail to reproduce the available observational constraints (for review, see \citealp{SG04}).  
With the upcoming Juno mission, Jupiter will not only be back in focus, but the new data may also 
invalidate models that cannot cope with the expected accurate measurements for the 
gravity field, the moment of inertia, or the oxygen abundance in the atmosphere.  

As a major source of influence on Jupiter models, \citet{GZ99a} identified the 
distribution of metals and helium onto certain layers. \citet{SG04} demonstrated
in addition the sensitivity of the Jupiter models to the equation of state (EOS) of hydrogen.
In the past few years, the H EOS in the warm dense matter region where
hydrogen metallizes has been improved by applying ab initio simulations
\citep{Bonev+04,Vorberger+07,Holst+08,Lorenzen+10,Morales+10a,Morales+10b,TamBonev10,Caillabet+11}.
The corresponding Jupiter models were consistent with the available observational constraints
(\citealp{Nettelmann+08}, hereafter N08). However, the measured enrichment of Jupiter's atmosphere 
with noble gases ($\sim 2\times\:$solar), and with carbon and nitrogen ($\sim 4\times\:$solar) 
suggests that oxygen is equally abundant, leading to an atmosphere metallicity of 2--4$\times\:$solar. 
Such high values could not be obtained with the former H EOS (H-REOS.1) used in N08, which 
instead required solar metallicity atmospheres. This low atmospheric metallicity was confirmed 
by another group that used the same method of computing the high-pressure EOS data \citep{Militzer+08}. 
Apart from this similarity however, the Jupiter models suggested by these two groups
differed from each other substantially with respect to core mass, heavy element content, 
and internal layering. In this paper we present new Jupiter models with $\sim 2\times\:$solar 
metallicity atmospheres as a result of applying an improved ab initio H EOS. 

In \sect\ref{sec:meth} we compare the improved H EOS (H-REOS.2) with H-REOS.1 and describe our 
procedure of structure and evolution modeling.
In \sect\ref{sec:res} we present the core mass and metallicity of our new Jupiter models and show 
that the difference in the Jupiter models of \citet{Militzer+08} and our group 
does not originate from the assumption of two or three layers (\sect\ref{ssec:res2L}).
We also present calculations of the moment of inertia (\sect\ref{ssec:nmoi}), of the 
characteristic frequency $\nu_0$ of the global oscillations (\sect\ref{ssec:nue0}), 
and of the thermal evolution (\sect\ref{ssec:cool}). We discuss Jupiter's atmospheric metallicity
in \sect\ref{sec:diss} and give our conclusions in \sect\ref{sec:concl}. Appended to this
paper are descriptions of our method to calculate the  entropy (Appendix \ref{apx:entropy}) 
and the moment of inertia (Appendix \ref{apx:moi}).

\section{Methods}\label{sec:meth}

\subsection{Improved H EOS}\label{ssec:HEOS}

Our new hydrogen equation of state (H-REOS.2) is constructed in the same way as the
former version H-REOS.1 (N08). The H EOS is assembled from different contributions
at the low density range ($\rho<0.1$ {\gccm}) and the high density range 
($0.2 \leq \rho \leq 9$ {\gccm}). 
At low densities, pressure dissociation and pressure ionization do not play an important 
role for the EOS. For these densities, models within the chemical picture agree 
well with experimental data. As for H-REOS.1, we here apply the Fluid Variational Theory 
(FVT$^+$) model \citep{Holst+07} where ionization is neglected for temperatures below 3000~K. 

In the high density range, our EOS table is obtained from finite-temperature 
density-functional theory molecular dynamics (FT-DFT-MD) simulations using the code VASP 
\citep{Kresse1993,Kresse1994,Kresse1996}. There are three improvements of our new 
H EOS. First, due to the enormous increase in computing power we were able to perform 
the simulations with 256 particles in a box compared to 64 particles \citep{Holst+08}. 
Second, we performed them up to 50,000~K, in order to cover the interior of a warm, 
young Jupiter, compared to 20,000~K (\citealt{Holst+08}, N08). In the simulations, a 
quantum mechanical treatment of the molecular vibrations of the ions is not included. 
Thus thirdly, after completion of the simulations, we add the energy of a quantum mechanical 
oscillator, $u_{vib}=k_{\mathrm{B}}\Theta_{vib}(0.5+(\mathrm{exp}(\Theta_{vib}/T)-1)^{-1})$,
to the internal energy of each molecule and subtract its classical energy of vibration, 
$k_\mathrm{B} T$. The degree of dissociation is estimated via the coordination number 
(see \citealp{Holst+08}). This contribution is non-negligible if the temperature of the system is 
below the vibration temperature $\Theta_{vib}$ \citep{FR09}. For hydrogen, we use the experimental
value $\Theta_{vib}=6338.2$~K. 

\begin{figure}
\plotone{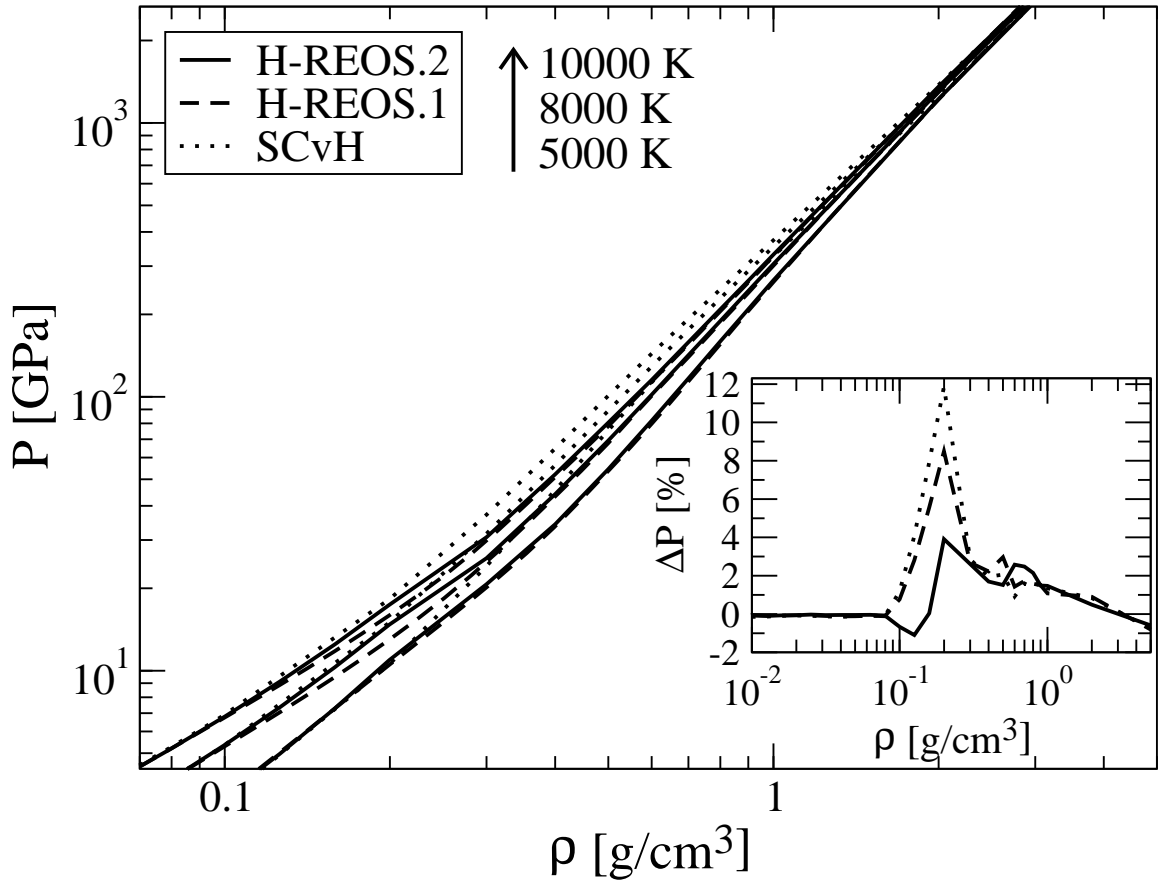}
\caption{\label{fg:isoT}Hydrogen isotherms of different EOS; \emph{solid:} H-REOS.2,
\emph{dashed:} H-REOS.1, and \emph{dotted:} SCvH-i EOS. Inset: The relative deviation of H-REOS.2 
with respect to H-REOS.1 for selected isotherms (5000 K: \emph{solid}, 8000~K: \emph{dotted}, 
10000~K: \emph{dashed}.)}
\end{figure}

\begin{figure}
\plotone{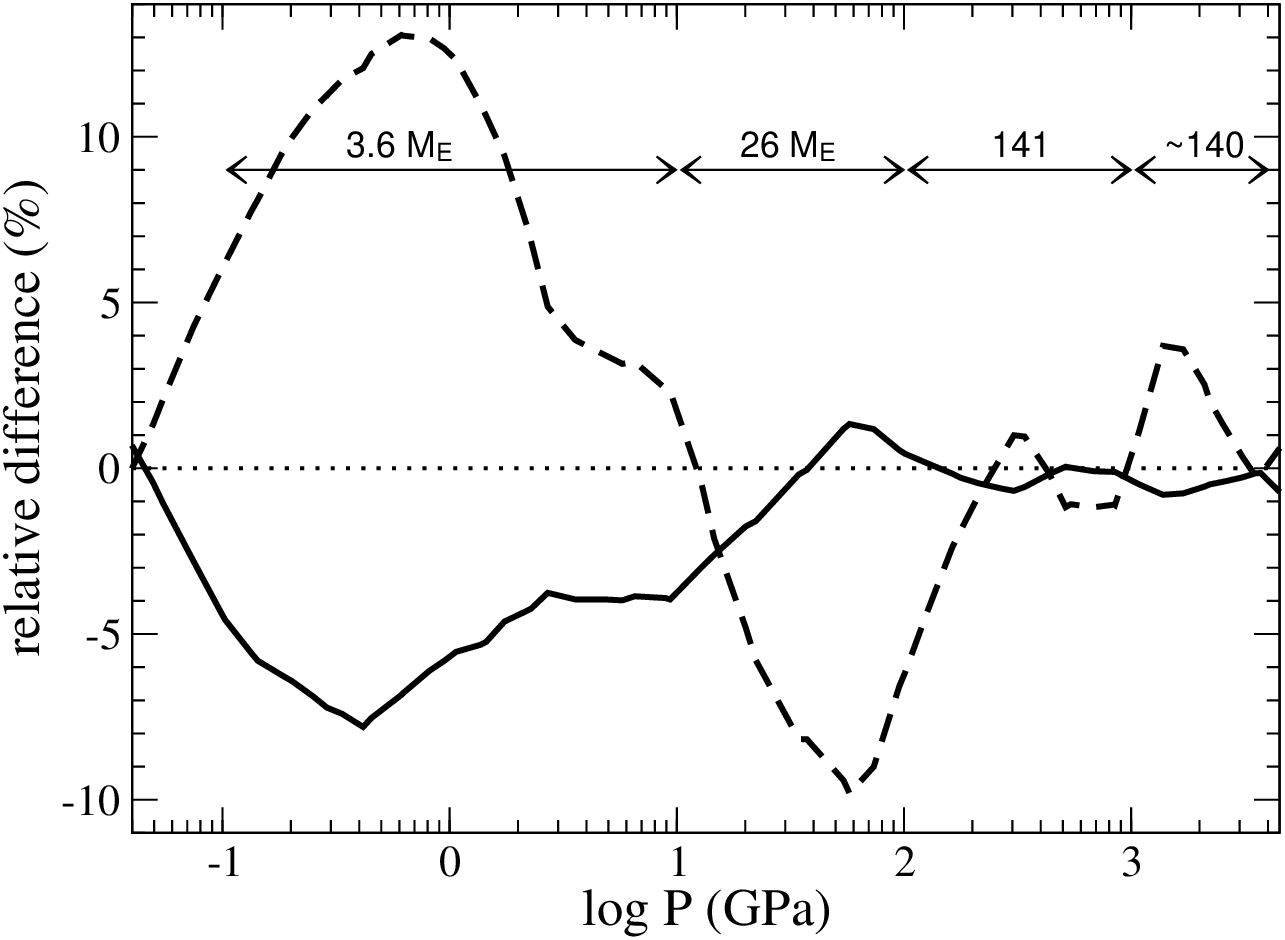}
\caption{\label{fg:isen}
Relative difference in density (\emph{solid}) and temperature (\emph{dashed}) 
between the Jupiter adiabats calculated with H-REOS.1 and H-REOS.2, for a H-He mixture with 
$Y=0.275$ (Jupiter average). The improved simulation of hydrogen dissociation causes systematically 
lower densities in the outer $\sim 16\:\ME$ of Jupiter. Numbers denote the mass in Earth masses between 
the pressure levels (in GPa) of 0.1-10, 10-100, 10-1000 as indicated by the arrows, and further 
up to 40 Mbar close to the core mantle boundary.} 
\end{figure}

For H-REOS.1 we obtained convergence of the EOS data within 3\% and estimated the overall uncertainty 
due to both statistical and systematic errors to be below 5\%. The new ab initio data in H-REOS.2 have 
an uncertainty of 1\% for most of the density-temperature space. They are within the uncertainty of the 
former data for $\rho\geq 0.3$ {\gccm} but shifted to 2\% higher pressures in a $T$--$\rho$ region where 
pressure dissociation and ionization occurs. Figure \ref{fg:isoT} shows selected isotherms. 
With the onset of pressure dissociation at 0.2~{\gccm}, the difference rises to 10\%. 
This is a rather small density for FT-DFT-MD simulations that requires large computational 
effort to reach converged results.
The displayed region of 5000--10000~K and 0.2--1~{\gccm} is relevant for the Jupiter adiabat.
Here, the  SCvH-i hydrogen EOS \citep{SCvH95} shows systematically higher pressures.
At densities of $0.1\leq \rho \leq 0.2$~{\gccm}, the difference in pressure is a consequence of 
a revised interpolation between the low density regime and the data points at 0.2~{\gccm}. 
This difference is particularly pronounced for the internal energy (not shown). Along the Jupiter
adiabat (\fig\ref{fg:isen}), the revised internal energy requires higher temperatures in order
to keep the entropy constant at the level defined by Jupiter's outer boundary condition of 170~K
at 1 bar (see \sect\ref{ssec:modeling}), and also leads to a decrease in the density at low
pressures of 0.1 to 10 GPa. However, due to little mass prevalent at those pressures in Jupiter, 
this region does not affect our Jupiter interior models. Instead, it is the 10--50 GPa pressure 
region where the lower densities of H-REOS.2 map onto the new Jupiter models.

Our EOS data tables provide the thermal EOS $P(\rho,T)$ and the caloric EOS $u(\rho,T)$, 
but not the specific entropy $s(\rho,T)$. In Appendix~\ref{apx:entropy} we describe our method 
to derive an adiabat through a given point $(P,T)$.


\subsection{Code improvements}

Careful inspection of our code used for the calculation of Jupiter's shape and rotation in N08 revealed 
that two coefficients, (one of second-order, the other one of third-order) contained errors.
Their effect on Jupiter models with H-REOS.1 is twofold; $i)$ the layer boundary 
shifts from $\sim$~8 to $\sim$ 4 Mbar when the outer envelope is made to have 
solar metallicity; $ii)$ the metallicity increases by $\sim 0.02$ (in absolute value)
in the outer envelope and $\sim 0.035$ in the inner one so that the maximum total mass 
of metals increases from 32 to $41\:\ME$, where the latter value
was presented in N08. All other results are almost unchanged. 
The two errors were found by a comparison of the resulting gravitational coefficients $J_{2n}$, 
$n=1$--4, of linear and polytropic density Jupiter and Saturn models with analytic solutions given 
in \citet{ZT78}, and with numerical solutions for these density distributions by T.~Guillot 
(pers.~comm.~2008). For details, see \citet{Nettelmann09}.
Another improvement of the code is the application of a Newton-Raphson scheme to find the central 
conditions that meet the envelope pressure and mass at the core-mantle boundary.
For the current version of our code, we here introduce the label MOGROP-11 
(MOdellierungsprogramm f{\"u}r GROsse Planeten, 2011.)

\subsection{Interior structure and evolution modeling}\label{ssec:modeling}

Apart from the code improvements mentioned above, our Jupiter interior and thermal evolution models
are generated by essentially the same code as introduced in N08, \citet{Nettelmann11}, and \citet{FN10},
where in the latter case the code was applied to the structure and evolution calculations of
Uranus and Neptune. We here recall the properties 
of our three-layer structure model for giant planets, review the observational constraints 
for the new Jupiter models, and describe the method of structure and evolution modeling.

\paragraph{Three-layer structure model.}
A sufficient number of parameters to meet the observational constraints and boundary conditions
are available if we assume a three-layer structure with two envelopes and
a core. The envelopes are adiabatic, homogeneous, and consist of hydrogen, helium, and 
heavy elements, while the core is made of rocks. Heavy elements in the envelopes
are represented by the water equation of state H$_2$O-REOS (\citealp{French+09}; N08), 
for helium we use He-REOS (\citealp{Kietzmann+07}; N08),
and for rocks the formula given in \citet{HM89}.  
The outer envelope has a lower He abundance ($Y_1$) than the inner envelope ($Y_2$).
We allow for different heavy element mass fractions in the outer $(Z_1)$ 
and in the inner envelope ($Z_2$). The transition pressure between the envelopes $P_{1-2}$
is a free parameter. The density and entropy are discontinuous at layer boundaries.

\paragraph{Observational constraints.}
The models are required to meet the observational constraints for the total mass $\MJ$, 
equatorial radius at the 1-bar pressure level $\RJ$, 1-bar temperature $T_1$, 
solid-body period of rotation $\omega$, gravitational moments $J_2,J_4,J_6$, 
atmospheric helium abundance $Y_{\rm atm}$, and have the same He mass 
fraction\footnote{Throughout this paper, the label $Y$ denotes a He mass fraction with 
respect to the H-He subsystem.} $Y$ as the protosolar cloud once had.
As long as the oxygen abundance below Jupiter's water cloud deck is not measured,
only the lower limit for the heavy element abundance $Z_{\rm atm}$ in Jupiter's atmosphere 
is known. This value is $1\times\Zsol$ (see \sect\ref{sec:diss}), where $\Zsol$ is the 
protosolar cloud's metallicity of 1.5\% \citep{Lodders03}.
  
These observational constraints for the structure models can be divided into four 
groups depending on whether they enter the modeling procedure explicitly or need to 
be fitted by adjusting model parameters, and whether or not they are varied within 
the 1$\sigma$ error bars. 
The first set (explicit, varied) is empty. The second set (explicit, not varied) consists 
of $\MJ=317.8338\:\ME$, $2\pi/\omega=$9h~55m~30s (the period of the magnetic field), the 
upper limit $T_1=170$~K, and $Y_{\rm atm}=0.238$. While the small uncertainties in $\MJ$ 
and $\omega$ would not affect the resulting models, lowering $T_1$ by 5~K would reduce $Z_1$ 
by $\sim 0.3\:\Zsol$ (N08), in the opposite direction to what we are aiming for. 
Modifying  $Y_{\rm atm}$ by $1\sigma$ ($\pm 0.007$) changes $Z_1$ by about $\mp 0.3\:\Zsol$. 
Because $Y_{\rm atm}$ was measured by the Galileo entry probe at  a depth where vertical convection 
ensures homogeneous distribution of those elements that do not form clouds (such as the noble gases) 
deeper inside, we set $Y_1=Y_{\rm atm}$.
The parameters of the third group (to be fitted, not varied) are $\RJ=7.1492\times 10^{7}$m 
and $Y=0.275$, whereof $\RJ$ is fitted by choosing the mean radius $\Rmean$ of the 
equipotential surface that coincides with the 1-bar pressure level, and $Y$ by adjusting $Y_2$. 
We find $\Rmean=10.955\:\RE$ and $Y_2=0.285$--0.325.
The gravitational moments $J_2$, $J_4$, and $J_6$ build the fourth group (to be fitted, varied). 
In particular, $Z_2$ is used to reproduce $J_2/10^{-6}$ within the tight observational bounds 
14697(1), and $Z_1$ is used to adjust $J_4/10^{-6}$ to either $-587$ (the observed central 
value) or $-589$ (the observed upper limit) \citep{Jacobson03}. When $J_2$ and $J_4$ are matched, 
$J_6/10^{-6}$ is found to lie within 34.0 and 37.5, consistent with the observed limits 
(29.1--39.5; \citealp{Jacobson03}). Our computation of Jupiter's thermal evolution makes use of 
the observed effective temperature $\Teff=124.4\pm 0.3$~K.

\paragraph{Modeling procedures.}
Internal profiles of $m$, $P$, $T$, and $\rho$ which depend on the radial coordinate $l$
are obtained by numerically integrating the structure equations 
\begin{equation}\label{eq:dPdl}
	\rho^{-1}\:dP/dl=-d(V+Q)/dl
\end{equation}
and 
\begin{equation}\label{eq:dmdl}
dm/dl = 4\pi\: l^2\:\rho\quad.
\end{equation}
Equation (\ref{eq:dPdl}) states that the force acting on a mass element with density $\rho$ 
at radius $l$ is balanced by a pressure gradient, where the force $-d(V+Q)/dl$ arises 
from the gravity field $V$ and the centrifugal force with potential $Q$ due to rigid-body 
rotation. Equation (\ref{eq:dmdl}) expresses the amount of mass $dm$ included in a spherical 
shell of density $\rho$ and thickness $dl$ at radius $l$. These two ordinary, first-order 
differential equations are integrated inward starting at the surface $l=\Rmean$ with the two 
boundary conditions $P(\Rmean)=1$~bar and $m(\Rmean)=\MJ$. In order to satisfy in addition the 
inner boundary condition $m(0)=0$, a further parameter is needed. For this we choose the core 
mass $\Mcore$. Thus for given values of the observational constraints and of the transition 
pressure $\Ptrans$, the result is one single Jupiter model, obtained in terms of the 
internal profile and values for $Z_1$, $Z_2$, and $\Mcore$. To calculate the planetary shape 
and the gravitational moments we apply the \emph{Theory of figures} \citep{ZT78} up to third 
order as in N08 and also to fourth order. Such structures describe Jupiter at present 
time $t_0=4.56$~Gyr.

For modeling Jupiter's thermal evolution we apply a procedure similar to that of \citet{Saumon+92}. 
To describe Jupiter's interior at earlier times $t<t_0$, we select a few representative 
structure models and generate for each of them a sequence of $\sim 80$ warmer interior profiles 
by increasing $T_1$ up to 800~K while $m(\Ptrans)$, 
$\Mcore$, and the element abundances are kept at constant values. Further models with 
intermediate $T_1$ values are then generated by interpolation. 
While for the structure models at present time with known $\RJ(t_0)$ the core mass was 
chosen to ensure mass conservation, we here invert the problem to find the planet radius $\Rmean(t)$ 
for given core mass. Unlike \citet{Saumon+92}, rotation is included in the approximation of 
spherical symmetry where the centrifugal force reduces to the zero-order term $-dQ/dl=(2/3)\omega^2 l$. 
We calculate the cooling curves in the usual approximation of constant angular velocity
\citep{SG04,Fortney+11}, and then also by including angular momentum conservation \citep{Hubbard70},
which requires just 1--3 additional iterations for a given value of $T_1$, and 
the corresponding change in the energy of rotation, $dE_{\rm rot}$.
The time $dt$ passed between profiles with different surface temperatures and hence different 
internal entropies is given by the equation of energy balance,
\begin{equation}\label{eq:LLL}
	\Leff - \Leq = -d\Eint/dt\quad,
\end{equation}
where $\Leff$ is the planet's observable luminosity in the infrared and $\Leq$ the luminosity 
the planet would have in equilibrium with the insolation. Using the \emph{Stefan-Boltzmann law} 
for the energy emitted by a black body with uniform temperature, these luminosities can be 
used to define the commonly used temperatures $\Teff$, $\Teq$, and $\Tint$. This way,
$\Teq(t_0)$ is derived from the $\Leq$ value given in \citet{Guillot05}. It is set constant over 
time or, alternatively, varied with time according to a linear increase of the irradiation flux 
$F_{\rm eq}:=\Leq/4\pi \Rmean^2$, starting with $F_{\rm eq}(0)=0.7 F_{\rm eq}(t_0)$ as predicted by
theoretical standard models for the Sun \citep{Bahcall+95,BS03b}. Note that the 
solar standard evolution model predicts an increasing bolometric luminosity with time, whereas 
the short-wavelengths (chromospheric and coronal) emissions of the young Sun may have been up to 
1000 times stronger than those of the present Sun as indicated by the measured XUV fluxes of young 
solar-like stars \citep{Ribas+05}. Absorbed XUV irradiation can lead to heating of and mass loss
from the upper planetary atmosphere \citep{Lammer+03} but absorption at high altitudes is not suspected 
to influence the energy balance of the interior \citep{GuillotShow02}. Therefore, we ignore the 
activity-related, time-dependent solar XUV flux in our evolution calculations for Jupiter.
 
The difference between the luminosities radiated away and absorbed equals the loss of the planet's 
intrinsic energy $d\Eint$ per time, see Eq.~(\ref{eq:LLL}). In the envelope, the energy 
lost per mass shell simply is the heat $\delta q=T(m)\:ds(m)$. 
\citet{Guillot+95} offer a convenient closure relation by relating $\Teff$ to $T_1$, 
namely $T_1=K\:\Teff^{1.244}\:g^{-0.167}$ according to the model atmosphere grid by 
\citet{Graboske+75}, where the parameter $K$ can be used to reproduce the observed $\Teff(t_0)$, 
and $g$ is the surface gravity. Also like \citet{Guillot+95}, we take into account the 
core's contribution to the planet's intrinsic luminosity.
The heat loss of the core is $\Mcore\:c_v\:dT_{\rm core}$, where we use $c_v=1\rm\:J\,K^{-1}\,g^{-1}$
as in \citet{Guillot+95}, and $dT_{\rm core}$ is the temperature difference of the isothermal
cores of subsequent interior profiles with different $T_1$ values, as is $ds(m)$ the 
entropy difference at mass shell $m$ between those interior profiles. Although the abundances
of radioactive elements decrease exponentially with time so that the energy production 
from radioactive decay in rock with meteoric composition may have been an order of magnitude larger 
in the past than it is today, we here use the present Earth's value $L_{\rm radio}=2\times 10^{13}$
$\rm J\,s^{-1}\,\ME^{-1}$ to account for this contribution. This treatment seems justified
for Jupiter because we find the prolongation of the cooling time due to the core's heat loss 
to be 0.01~Gyr only. Thus the cooling equation  reads 
\begin{equation}\label{eq:cool}
dt = -\frac{\int_{\Mcore}^{\MJ}dm\:T\,ds + \Mcore c_v dT_{\rm core} + dE_{\rm rot}}
           {4\pi R^2\sigma (\Teff^4-\Teq^4) - L_{\rm radio}}\quad. 
\end{equation}
After integrating Eq.~(\ref{eq:cool}) backward in time we obtain the cooling time $\tau$ that 
the particular Jupiter model needs to cool down from an arbitrarily hot initial state to the
present state.

\section{Results}\label{sec:res}

\subsection{Core mass and metallicity of three-layer models}\label{ssec:res3L}

\begin{figure}
\epsscale{.65}
\plotone{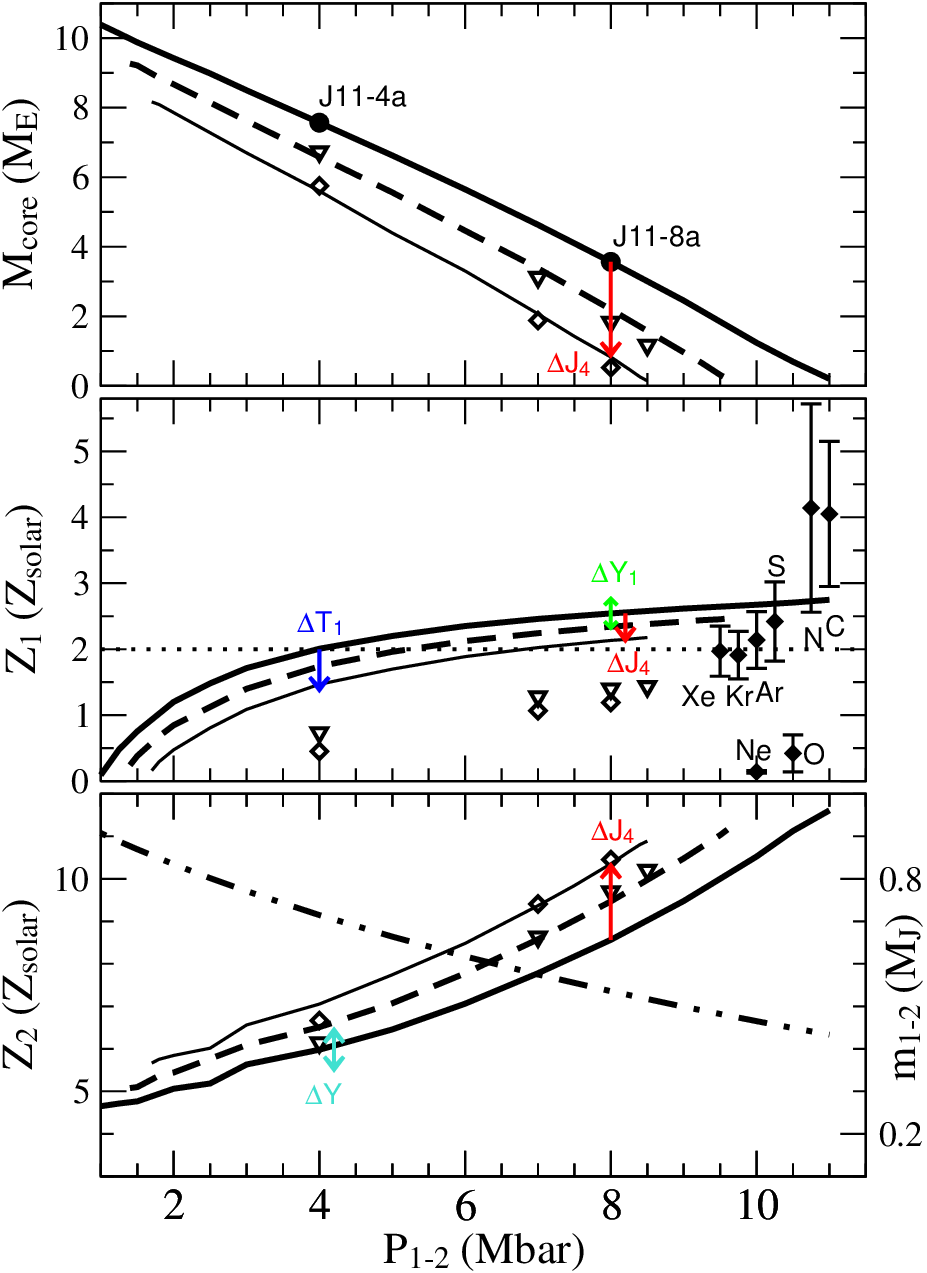}
\caption{\label{fg:McZZP12}
Jupiter interior models for different transition pressures $P_{1-2}$ at the layer boundary. 
\emph{Top:} core mass; \emph{middle}: outer envelope metallicity scaled to the solar value 
$Z_{\rm solar}=0.015$; \emph{bottom}: inner envelope metallicity. 
\emph{Curves} are with the improved H EOS (H-REOS.2) and \emph{open symbols} are models 
calculated with our former H-REOS.1. Among these, \emph{solid curves} and \emph{triangles} 
are for $J_4/10^{-4}=-5.89$, while \emph{dashed curves} and \emph{diamonds} are for $J_4/10^{-4}=-5.87$. 
The \emph{thick} (\emph{thin}) curves are calculated with the Theory of Figures to 3rd (4th) order. 
Arrows indicate the shift of the models J11-4a and J11-8a (\emph{filled circles}) if the observational 
$1\sigma$ error bars of $Y$, $Y_1$, $T_1$, and $J_4$ are applied as described in \sect\ref{ssec:res3L}.
\emph{Filled diamonds} are measured atmospheric abundances in solar units 
placed arbitrarily on the $x$-axis. The \emph{dotted} line is a guide to the eye for a (minimum)
outer envelope metallicity of $2\times$ solar. The new models with H-REOS.2 are consistent
with this constraint, but H-REOS.1 based models were not. The \emph{dot-dot-dash line} in the 
bottom panel shows the scaled mass coordinate $m_{1-2}:=m(\Ptrans)$, which is within the line thickness 
independent on the model assumptions. It can be used to calculate the mass of heavy elements in the envelopes
$M_{Z,\rm env} = Z_1\times (\MJ-m_{\rm 1-2}) + Z_2\times(m_{\rm 1-2}-\Mcore)$ .}
\end{figure}

Results for the core mass, the outer envelope metallicity, and the inner envelope metallicity
as functions of the depth of the layer boundary between the envelopes are shown in \fig\ref{fg:McZZP12}.
If the layer boundary is located higher in the planet (lower transition pressures),
$Z_1$ and $Z_2$ become smaller. The lighter envelopes then require a larger core 
mass to conserve the total mass. When $Z_1$ decreases down to zero, the core mass
cannot grow any further. However, we do not consider models with $Z_1<\Zsol$ acceptable
as this is inconsistent with Jupiter's observed atmospheric abundances.
We find $Z_2\gg Z_1$ and $\Mcore < 10\:\ME$ for all our acceptable three-layer models.
As the gravitational moments are most sensitive to the density in the outer part 
of the planet at pressures of a few Mbars ($J_2$) or even below 1 Mbar ($J_4$) and increase
with the local density, $Z_1$ must decrease when the denser inner envelope extends 
farther out. Otherwise, $|J_4|$ would become too large. 
On the other hand, as we go to higher pressures deeper inside the planet, the sensitivity of $J_4$, 
and later also of $J_2$ approaches zero. Therefore, $Z_1$ changes only weakly with $\Ptrans$
for deep internal layer boundaries. However, since $J_2$ is adjusted by $Z_2$, $Z_2$ must rise
strongly with $\Ptrans$ in order to provide a sufficiently high mass density that is able to
keep $J_2$ on the high level of the observed value. With rising metallicity the
envelope becomes denser, so the core mass must decrease. When $\Mcore=0$ is reached, the layer
boundary is as deep as it can be, and $Z_1$ adopts a maximum value. For models
with H-REOS.1, this maximum is $1.0\:\Zsol$  if $J_4/10^{-6}$ is adjusted to 587, 
and $1.5\:\Zsol$ if $J_4/10^{-6}$ is adjusted to 589. For models with our improved H EOS, this
maximum is $2.5\:\Zsol$ ($J_4/10^{-6}=587$) and $2.7\:\Zsol$ ($J_4/10^{-6}=589$), respectively.
This enhancement in $Z_1$ arises from the $\sim 2\%$ higher pressures at given hydrogen 
mass density (\fig\ref{fg:isoT}), implying $\sim 2\%$ lower hydrogen mass density at given
pressure. The lower partial density of hydrogen in the mixture requires to be compensated for
by a similar amount of metals.

Including 4th order coefficients to the calculations of Jupiter's shape and rotation systematically
shifts the solutions to $\sim 0.8\:\Zsol$ lower outer envelope metallicities, inducing an
additional uncertainty on $Z_1$ that is of the same size as the error bars of single observables 
such as $T_1$ and $J_4$, see below. For $\Mcore$ and $Z_2$, the inclusion of higher-order terms has 
an even bigger effect than the improvement in the H EOS. Obviously, 
a convergence check of the solution with increasing order of the theory is highly important 
but beyond the scope of this paper. The accuracy of our current computer code is insufficient 
to reliably study the effect of higher than 4th order coefficients. From preliminary calculations,
solutions with 5th order coefficients are found to lie between those of 3rd and 4th order.


Figure~\ref{fg:McZZP12} also shows the observed element abundances in Jupiter's atmosphere.
Assuming that O is equally enriched as C, these abundances indicate an average enrichment
of 2 (noble gases) to 4 (C,N,O). Taking $2\times$ solar as the lower limit for $Z_1$, 
models with $\Ptrans < 4$~Mbar drop out of the realm of acceptable
models, reducing the upper limit for the core mass down to $8\:\ME$.
The total mass of metals is $M_Z=29$--$32\:\ME$.

As representatives for our acceptable models we recommend models J11-4a and J11-8a,
highlighted in \fig\ref{fg:McZZP12}. Model J11-4a is the one with the outermost possible 
layer boundary, i.e.~at 4~Mbar. Model J11-8a has $\Ptrans=8$~Mbar for which $Z_1$ is $\geq 2\Zsol$
for all of the above considered uncertainties. A machine-readable table for model J11-4a is provided 
as supplemental material, see Table~\ref{ascitab}.
For these two models we also vary $Y$, $Y_1$, $T_1$, and $J_4$ within their 1--$\sigma$ error bars.  
Modifying $Y$ by $\Delta Y=\pm 0.05$ influences $Y_2$ in the same direction, and thus leads to a 
change in $Z_2$ of $\mp 8\%$. Modifying $Y_1$ by $\Delta Y_1=\pm 0.05$ causes a response in $Z_1$
of $\mp 10\%$. Similarily, a 5~K colder 1-bar level leads to a 50\% decrease in $Z_1$. 
In both cases, $Z_2$ consequently changes by $\pm 3$--4\% in the opposite direction to $Z_1$, and 
$\Mcore$ changes by $\mp 2$--3\%, again opposite to $Z_2$. The colder outer envelope adiabat also cools
the interior, resulting in a core that is 100~K colder. The slightly denser H-He subsystem that
results reduces the amount of metals in the envelope required to match $J_2$, but the net effect 
of $\Delta T_1$ on $Z_2$ remains positive. Setting $|J_4|/10^{-6}$ down to 585 lowers $Z_1$ by $\sim 15\%$. 
Clearly, the response of $Z_2$ (increase) and thus of $\Mcore$ (decrease) in this case is stronger 
than in case of $\Delta T_1$ and $\Delta Y_1$.
Including these uncertainties, the total mass of metals changes but insignificantly to 28--32~$\ME$.

\subsection{Core mass and metallicity of two-layer models}\label{ssec:res2L}

\begin{figure}
\plotone{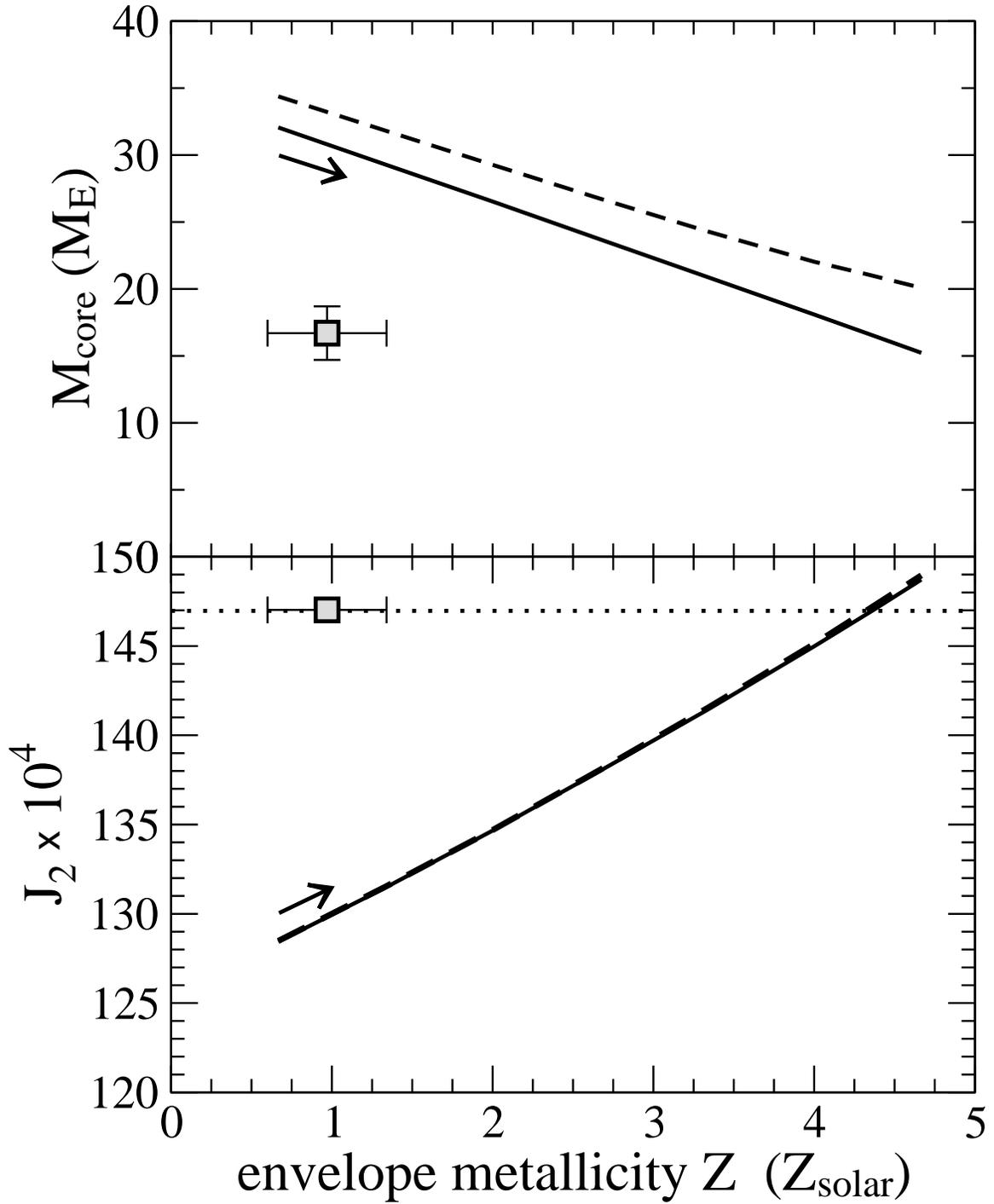}
\caption{\label{fg:zmc2L}
Two-layer models on their way to meet Jupiter's observed $J_2$ at $146.97\times 10^{-4}$ 
\emph{(vertical dotted line)}, starting at low envelope metallicity and moving in the direction 
as indicated by the arrows. The converged solution of our calculations is obtained through 
variation of $Z$. It has $Z=4.33\:\Zsol$ and a core mass of $21.1\:\ME$ if the core is mostly ice 
(\emph{dashed}), or $\Mcore=16.5\:\ME$ if made of rocks (\emph{solid}), respectively. 
\emph{Square with error bars}: the M08 Jupiter model.}
\end{figure}

The striking difference between the \citet{Militzer+08} (hereafter M08) predictions for the 
core mass of Jupiter ($16\pm 2\:\ME$) and that of our group (0--$8\:\ME$), where both groups applied 
similar ab initio simulations to generate the EOS data, has prompted speculations about 
causes for that difference \citep{MH09}.
We here investigate how this difference relates to the modeling assumptions of \emph{two}-layer 
(M08) or \emph{three}-layer (N08, this work) structures. 
For that purpose we have calculated two-layer models following M08. As in M08, the homogeneous
envelope has $Y_1=0.238$, and the missing helium to obtain an average $Y$ of 0.275 is not 
explicitly accounted for but may be part of the core mass. The core is either pure rocks or rock-ice,
where we switch from ice to rock at 70 Mbar. As a result, only the innermost $4\:\ME$ are 
rocks. Because these models do not have the degree of freedom $Z_2$, we can only match either
$J_2$ or $J_4$. As in M08 we choose to match $J_2$ by varying the envelope metallicity $Z$.
We apply our usual numerical procedure (see \citealp{Nettelmann11} for details) 
and vary Z, starting from 0.01, until $J_2(Z)$ meets the observed value. The core mass must 
decrease with increasing envelope metallicity (denser envelopes) in order to ensure total mass 
conservation. If the number of layers assumed for the structure of the model were responsible for 
the different Jupiter models we would expect to obtain the same two-layer model as in M08. 
This is \emph{not} the case, as illustrated in \fig\ref{fg:zmc2L}.
While the M08 models require $4\pm 2\:\ME$ heavy elements in the envelope to meet $J_2$, 
corresponding to  $Z=1.0\pm 0.5\:\Zsol$ for a $16\:\ME$ core, our two-layer models require  $Z=4.3\:\Zsol$.  

This significantly larger envelope metallicity of our two-layer models enables us to make 
a transition to three-layer models where metals are shuffled from the outer part of the envelope 
to the inner part in order get $|J_4|/10^{-6}$ down to the observed value (from 611 to 587). 
The M08 models do not have this degree of freedom as the already low envelope metallicity 
would become zero long before $J_4$ is matched.
Hence the M08 models stay at $J_4/10^{-6}=614$ and a large core while our two-layer models can 
undergo a transition to three-layer models where $Z_1$ becomes smaller and $Z_2$ larger than in 
the homogeneous envelope case. As a consequence of large $Z_2$ values, the core mass decreases further.  
In contrast, the core mass of the two-layer models is large because this mass also contains 
the missing mass of helium $dM_{\rm He}=(0.275$--$0.238)(\MJ-\Mcore)\sim 11\:\ME$ needed for an 
overall abundance $Y=0.275$. Thus the ice-rock core mass of two-layer models is about 5--10$\:\ME$, 
just at the upper limit of our three-layer models. This estimate would also hold for the two-layer
model of M08. However, we cannot reproduce their Jupiter model, 
the reason for which remains unexplained at the moment. 
We conclude that the difference between the M08 two-layer models and our three-layer models does
not originate from assumptions about the number of layers, but instead is a symptom of already 
different models within the two-layer frame.


\subsection{Moment of inertia}\label{ssec:nmoi}

We have calculated the moment of inertia $I$ as described in Appendix~\ref{apx:moi}. The non-dimensional
form (\nI) is obtained as ${\nI}=I\,\MJ^{-1}\,\Rmean^{-2}$. For all of our three-layer models, 
we find an almost invariant value $\nI=0.27605\pm 0.03\%$. If we ignore Jupiter's shape deformation, 
the resulting {\nI}  decreases by 4\% down to $0.26539\pm 0.03\%$. This is close to the predictions 
of 0.2629$-$0.2645 of \citet{Helled+11}, who considered a broad set of interior models with core 
masses between 0 and $40\:\ME$ as allowed by the use of six-order polynomials to represent the 
pressure-density relation in Jupiter's envelope.
From our calculations we conclude that {\nI} is not an appropriate parameter to constrain
Jupiter's core mass further. In contrast, \citet{Helled+11} found a significant variation of 0.6\% 
over the full range of their models and still a 0.2\% variation when excluding models with $\Mcore$
larger than $10\:\ME$. Juno's measurement of Jupiter's {\nI} is expected to have an accuracy of 0.2\%
\citep{Helled+11}, hence sufficient to distinguish between the different predictions.

\subsection{Characteristic frequency}\label{ssec:nue0}

Jovian seismology has been long recognized as a unique opportunity to infer interior structure
properties \citep{Vorontsov+76}. In particular, the predicted low-frequency acoustic free oscillations 
depend sensitively on the core mass and internal layer boundaries of theoretical Jupiter models
\citep{GZ99b}. However, the expected small amplitudes ($\sim 10$~cm) and Jupiter's rapid rotation 
make it difficult to measure acoustic modes. Recently, \citet{Gaulme+11} used the 
SYMPA spectrometer to detect the lowest frequency modes ($\sim 1000\:\mu$Hz) of Jupiter from spatially 
resolved radial velocity measurements. In the approximation of low degrees $l$ and overtones of high 
radial order $n\gg l$, the acoustic modes $\nu_{n,l}$ are proportional to the characteristic 
frequency $\nu_0$ (also called equidistance) and are equally spaced by $\nu_0$ for given $l$: 
$\nu_{n,l}\simeq (n+1/2)\nu_0$. \citet{Gaulme+11} observed $\nu_0=155.3 \pm 2.2\:\mu$Hz. 
The inverse of $\nu_0$ is twice the time a sound wave needs to travel from the center to the 
surface (Jupiter's troposphere),
\begin{equation}\label{eq:nue0}
	\nu_0=\left[2\int_0^{\RJ}dr\:c_s^{-1}\right]^{-1}\:, 
\end{equation}
a useful parameter that interior models can easily be compared to. In Eq.~\ref{eq:nue0},
$c_s$ is the adiabatic sound velocity.
\citet{GZ99b} calculate $\nu_0=152$--$155\:\mu$Hz for SCvH-i EOS based Jupiter models with five 
layers and core masses of 3--$10\:\ME$.
\citet{Gaulme+11} obtain the same range of values for $\nu_0$ with three-layer models of Jupiter 
that have a small internal He discontinuity and core masses of 0--$6\:\ME.$
For our three-layer models with $\Mcore=0$--$8\:\ME$ we find $\nu_0=155.7$--$156.3\:\mu$Hz. 
Since this range is rather narrow and slightly above that of former Jupiter models, a more accurate 
determination of the global mode spacing may help to rule out some Jupiter models but not 
constrain Jupiter's core mass further within the current uncertainty of $\sim 0$--$10\:\ME$.

\subsection{Cooling curves}\label{ssec:cool}

\begin{figure}
\plotone{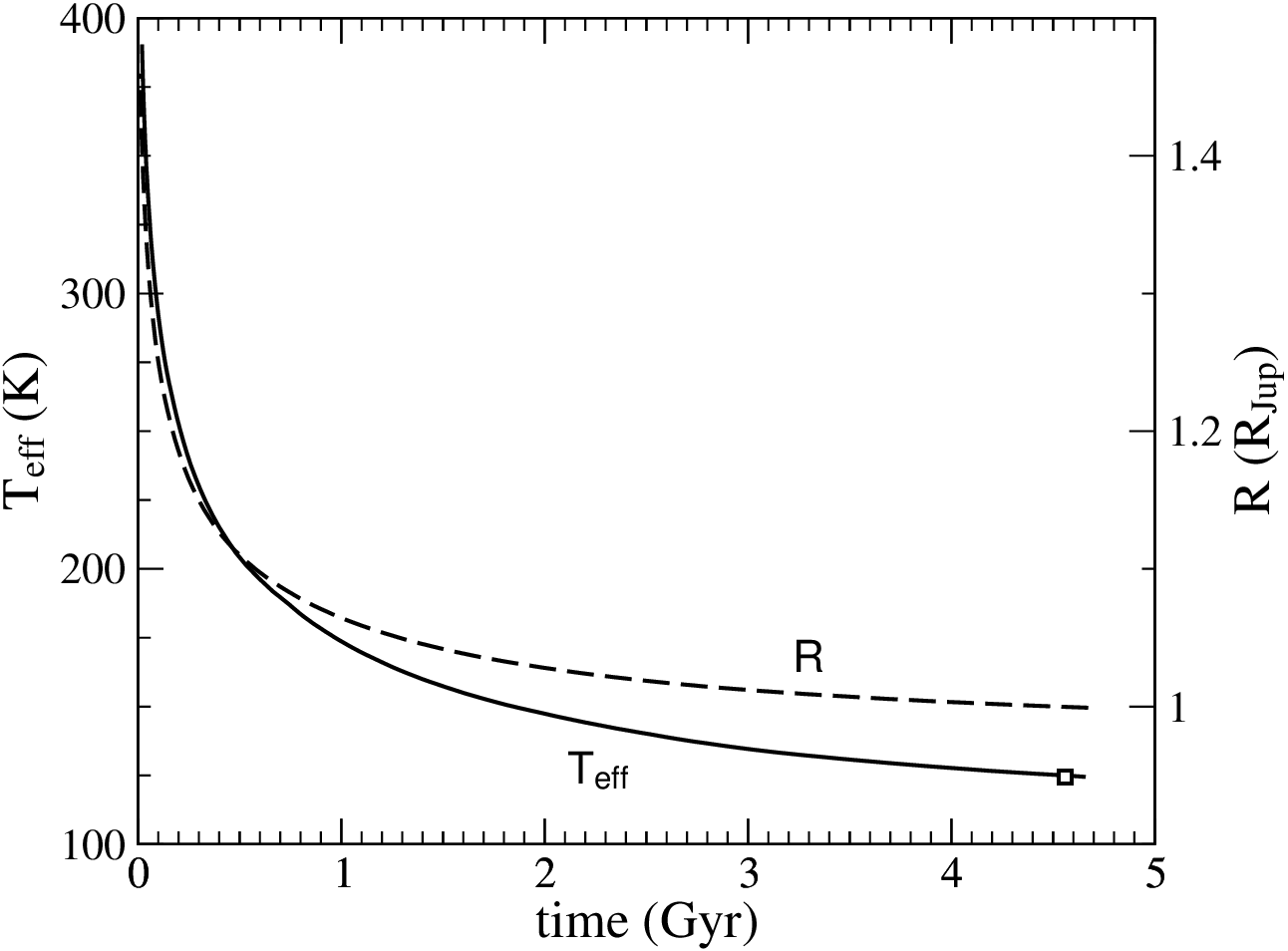}
\caption{\label{fg:evol}Effective temperature (\emph{solid}) and mean radius (\emph{dashed}) 
of the homogeneously cooling Jupiter (interior model J11-4a). 
The \emph{square} indicates the observed $T_{\rm eff}$ at present time, 
whose observational error bar of 0.3~K is not resolved in this figure. }
\end{figure}

For a better comparison  with published calculations of the cooling of Jupiter we first assume 
a constant irradiation flux and constant angular velocity over time. The latter assumption is justified 
by the rather small change of the rotational energy during evolution compared to Jupiter's intrinsic 
energy loss \citep{Hubbard77} and the small change in planet radius over most of the contraction 
time scale \citep{Guillot+95}. Our H-REOS.2 based Jupiter model J11-4a has a calculated cooling time 
of $4.66\pm 0.04$~Gyr, see \fig\ref{fg:evol}, and model J11-8a of $4.68\pm0.04$~Gyr,  where the 
uncertainty mostly arises from the observational error bar of $\Teff$. 
This result can be considered in good agreement with the age of the solar system 
($\tau_{\odot}=4.56$~Gyrs) given that none of the alternative proposed Jupiter models that rely on 
free-energy models for the equation of state \citep{SG04}, such as LM-H4 (4.0 Gyr), SCvH-i (4.7 Gyr), 
and LM-SOCP (4.8 Gyr) gives a better agreement. 
On the other hand, \cite{Fortney+11} investigated the uncertainty arising from the use of
the \citet{Graboske+75} model atmosphere grid for Jupiter's evolution. If instead a self-consistent 
model atmosphere grid was applied as it is the common approach for exoplanets, Jupiter's cooling 
time increases by $\sim 0.5$~Gyr. Sinking He droplets from H-He phase separation in 
Jupiter would also prolong the cooling time further. A process working in the opposite 
direction could be rising material from core erosion. 
During the first 500 Myr, Jupiter has shrunken from 1.4--1.5$\RJ$ down to 1.1$\RJ$ 
(\fig\ref{fg:evol}). Because the initial conditions for the long-term evolution of a $1\:\MJ$ 
planet are forgotten after 0.01 Gyr \citep{Marley+07}, we consider the displayed radius 
evolution at young ages realistic.

\begin{figure}
\plotone{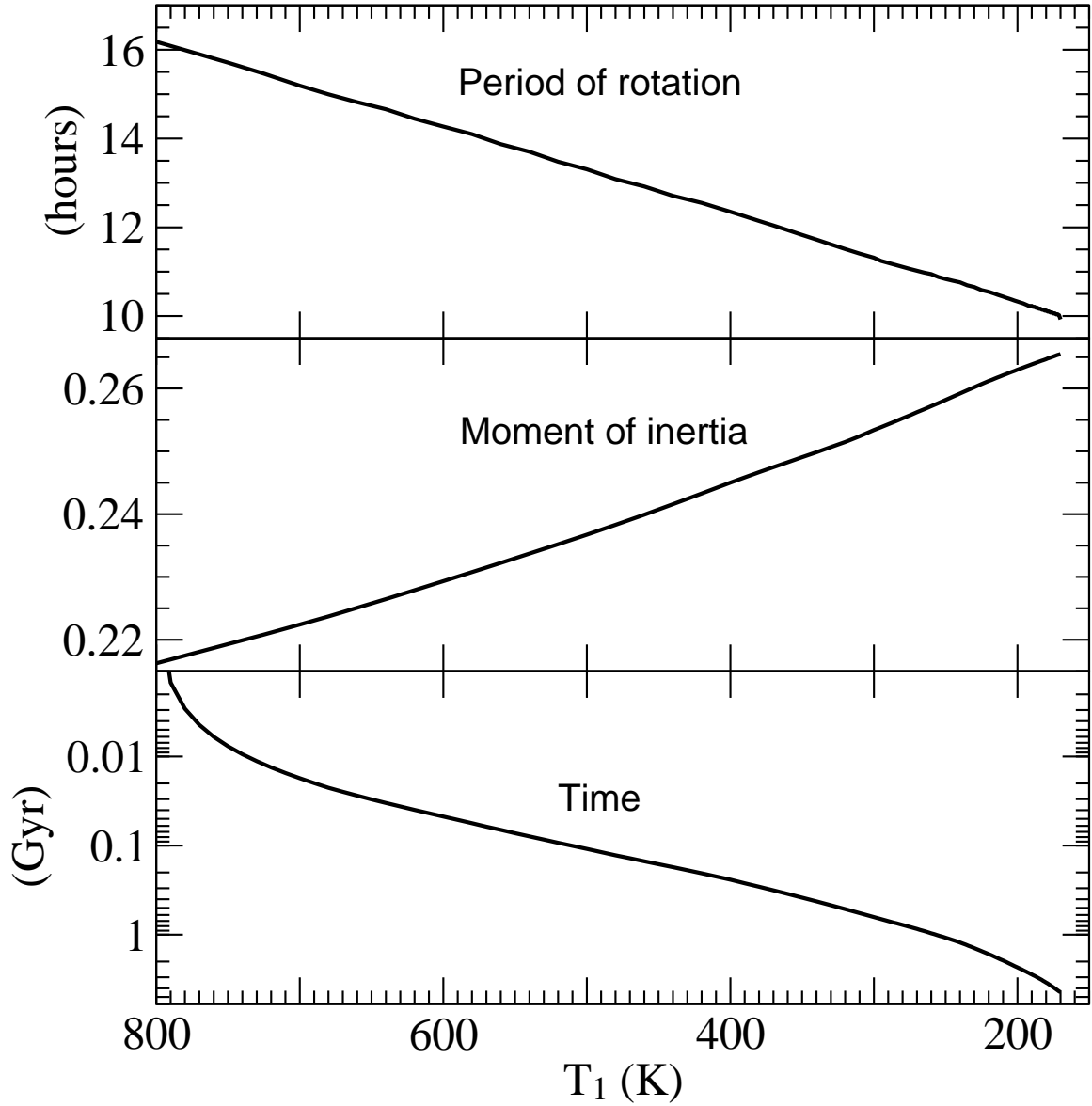}
\caption{\label{fg:evolTIP}
Jupiter's evolution including angular momentum conservation, the corresponding change in the energy of 
rotation, and solar irradiation that increases with time. \emph{Upper panel:} period of rotation; 
\emph{middle panel:} normalized moment of inertia {\nI} for the surface temperatures $T_1=170$--800~K 
that define the internal structure during the evolution. \emph{Lower panel}: Map of $T_1$ onto time.}  
\end{figure}

In a second step, we keep the angular momentum conserved to Jupiter's current value (within
a numerical accuracy of 0.4\% which gives sufficiently smooth evolution curves) and include the 
subsequent change of rotational energy $dE_{\rm rot}(t)$, as also done in \citet{Hubbard70}.  
Such second order effects are important for estimates of the size of the correction factors that are 
necessary to bring the cooling time in agreement with $\tau_{\odot}$.
It is clear that both effects ({\nI} and $dE_{\rm rot}$) will prolong the cooling time: $i)$ for the same 
total mass, a larger planet (the young Jupiter) rotates slower, implying a weaker centrifugal force that 
pushes matter outward, and the consequently smaller radius will allow less energy to be radiated away from 
the surface; $ii)$ angular momentum conservation ($L=I\omega$) then implies a lower energy of rotation 
$E_{\rm rot}=1/2\:L\omega$ at young ages. The increase in $E_{\rm rot}$ with time must be compensated for
by a reduced luminosity, implying again a longer cooling time. For model J11-4a, the first effect increases
$\tau$ by 0.1~Gyr, and the second one by 0.2~Gyr, so that we find $\tau=4.96\:$Gyr.
In a third step, we consider a time-dependent solar irradiation by a linear approximation of the Sun's 
luminosity evolution, which is assumed to have started with 70\% of the current value. We find the lower 
irradiation to speed up the cooling by $\sim 0.6$~Gyr, so that we end up with $\tau=4.41\pm 0.04$~Gyr. 
Figure \ref{fg:evolTIP} shows the relations between the structure parameter $T_1$ that determines the 
planet radius, $P=2\pi/\omega$ and {\nI}. Their evolution with time can be read from the cooling curve $T_1(t)$.

The resulting cooling times of 4.4~Gyr (linearly increasing insolation) to 5.0~Gyr (constant insolation) 
suggest a reasonable understanding of Jupiter's interior and evolution, where remaining uncertainties 
seem to  be attributable to uncertainties in the luminosity evolution of the Sun. 
This may be counter-intuitive as the present Sun is, by means of accurate helioseismology, 
far better constrained than Jupiter.  For instance, uncertainties in the sound velocity profile of 
the solar standard model are of the order of 0.1\% only \citep{BS03a}. On the other hand, 
non-standard solar evolution models that predict a bright and more massive young Sun \citep{BS03b} 
cannot completely be ruled out by observationally derived stellar mass loss rates, 
see \citet{Guedel07} for a review. 

However, homogeneous, adiabatic evolution models for Saturn and Uranus that are equivalent to 
those presented here for Jupiter keep failing to reproduce the observed luminosities \citep{Fortney+11}, 
indicating that the standard three-layer model assumption may be too simplistic for some giant planets. 
Therefore, \cite{LC12} investigated the possibility of layered convection in Jupiter, where heat is 
transported inefficiently by diffusion across thin stable layers, separated vertically by convective cells. 
They determined the maximum super adiabaticity that would give density distributions in agreement with 
the gravity field data. 
The deduced presence of roughly $10^4$ diffusive interfaces in Jupiter will qualitatively reduce 
the heat flux out of the deep interior and shorten the cooling time \citep{Stevenson85}. Quantitative 
estimates of (at least) this effect are necessary before one can conclude a reasonable understanding 
of Jupiter's evolution.


\section{Discussion}\label{sec:diss}

\begin{figure}
\plotone{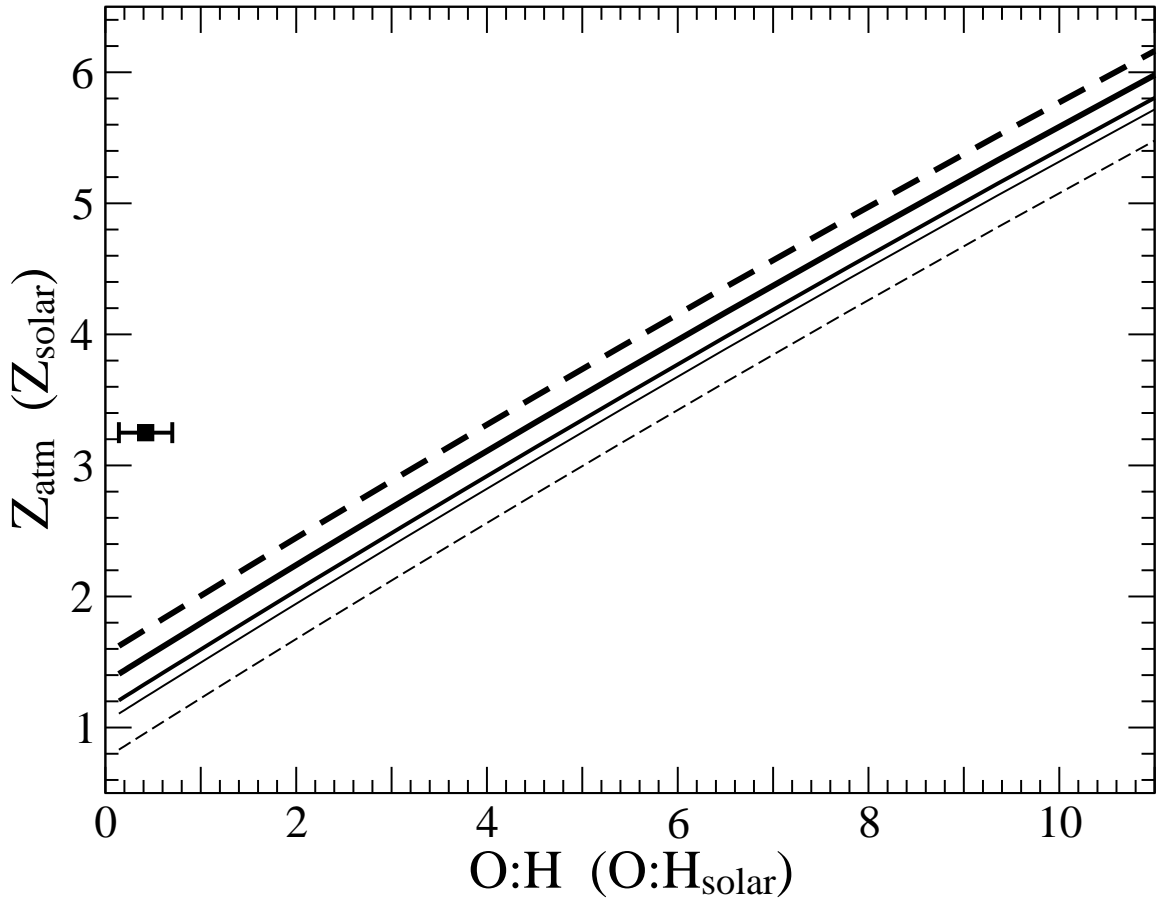}
\caption{\label{fg:ZvonOH}
Metallicity $Z_{\rm atm}$ in Jupiter's deep atmosphere and outer envelope in dependence on the O:H ratio
in solar units \citep{Lodders03}. \emph{Thin solid}: using the mean values of the measured
species (C,N,S,P,Ne,Ar,Xe,Kr); \emph{solid (thick solid)}: including in addition $1\times$ 
(3$\times$) solar abundances of (Mg,Al,Ca); \emph{thick dashed}: same as thick solid but using 
the upper limits of the measured abundances; \emph{thin dashed}: same as thin solid but 
using the lower limits. The measured O:H abundance at 19 bars \citep{Atreya+03} is 
indicated (vertical position has no meaning).} 
\end{figure}

According to our new Jupiter models with the improved H EOS (H-REOS.2), metals are enriched 
by a factor of at most 2.7 in the outer envelope and atmosphere. That value can be increased
to 3.0 if the atmospheric He abundance is lowered by $1\sigma$, but remains 
close to the lower limit of the measured C and N abundances. However,
Jupiter's atmospheric metallicity is not directly observable. What can be measured 
are the abundances of single chemical species from which the atomic ratios with respect 
to the number of H atoms can then be inferred. Such measurements have been obtained for Jupiter 
by remote sensing with the Voyager IR and UV spectrometers, IR spectrometry with the
Earth Orbiter ISO, Galileo orbiter measurements in the near IR, the Galileo Probe 
Mass Spectrometer (GPMS) data, and ground-based remote sensing from IR to radio 
wavelengths. Among the detected species, the measured abundances of C, N, S, P, Ar, Kr, and Xe 
(see \fig\ref{fg:McZZP12}) are taken to be representative for the convective region below
the 1-bar level, the outer boundary of our models, as they are not subject to non-equilibrium 
chemistry, cloud formation, or other processes that could significantly affect their abundance 
at such low altitudes \citep{Atreya+03}. Therefore, the enrichment of these elements is assumed 
to be representative also for those elements that are still subject to such processes, such as 
oxygen. The Juno orbiter, which is on route to Jupiter, is designed to measure 
the O abundance below the cloud level. In \fig\ref{fg:ZvonOH} we vary the O:H
ratio and calculate the corresponding metallicity of Jupiter's atmosphere with the help of 
Eq.~(\ref{eq:Z1}),
\begin{equation}\label{eq:Z1}
Z_1 = \frac{\sum_i\mu_i\,(\mbox{N$_i$:H})}{\mu_{\rm H} + \mu_{\rm He}(\mbox{He:H}) 
       + \sum_i\mu_i\,(\mbox{N$_i$:H})}\:,
\end{equation} 
where $\mu_i$ is the atomic weight of species $i$. By taking into account the measured 
abundances N$_i$:H with their error bars (mean, minimum, and maximum values) and including 
the non-detected elements \{Mg, Al, Ca\}, because they are abundant in the solar system, 
with enrichment factors of 0, 1, and 3$\times$~solar,  we aim to cover the uncertainty in
Jupiter's metallicity for a given O:H ratio. 
The unknown abundances of Mg, Al, and Ca contribute an uncertainty to Jupiter's $Z_1$ 
of $\pm 0.2\times \Zsol$, and the observational error bars of the measured species
contribute an additional $\pm 0.3\times \Zsol$.


Assuming Jupiter accreted its volatiles as a result of infall of icy planetesimals 
from cold, outer regions of the protosolar cloud which resembled the interstellar medium 
as we observe it today \citep{Owen95}, we would expect O:H to be similarly enriched in 
Jupiter as C:H \citep{Owen+99}, i.e.~O:H $\geq 3\times$ solar. For this ratio we derive 
from \fig\ref{fg:ZvonOH} a minimum atmospheric metallicity of $2\:\Zsol$ that Jupiter models 
should satisfy. A $\sim 2.5\times$ solar metallicity (this work) is consistent with an O:H 
of 2--4$\times$~solar. A measured O:H ratio greater than $4.5\times$ solar would not be 
consistent with the Jupiter models presented here. 
According to the uncertainties discussed above, the measured O:H ratio at the 19 bar level sets 
a lower limit to Jupiter's envelope metallicity of 0.8--$1.9\:\Zsol$ (based on \fig\ref{fg:ZvonOH}). 

The metallicity of the envelope is directly tied to the density of the H-He mixture for fixed $(T,P)$.
Our models are always warmer ($\Tcore\sim 20,000$~ K) than the M08 models 
($\Tcore\sim 14,000$~K) implying a less dense H-He mixture that allows to add more metals
compared to the M08 Jupiter adiabat. This is surprising as non-linear H-He mixing 
effects, which have been shown to decrease the density of the mixed phase \citep{Vorberger+07} 
compared to the linear mixture that we apply, are already included in the H-He EOS of 
the M08 Jupiter models.
Non-linear H-He mixing effects remain an appealing possibility 
(see N08) to enhance $Z_1$ in our three-layer Jupiter model calculations further. 

We favor the three-layer models over the two-layer models as they allow to reproduce $J_4$
without invoking the hypothesis of deep-zonal winds because the penetration depth of such
winds might have to be much deeper ($<0.96\:\RJ$) than physically allowed in the presence of 
convective flow motions \citep{Liu+08}. If deep-zonal winds indeed exist and require 
a correction of $J_4$ as proposed in M08, the atmospheric metallicity of our 
Jupiter models would then rise up to $4.5\:\Zsol$ and the ice-rock core mass up to $12\:\ME$.

The three-layer model framework is amenable to additional variations. \citet{FN10} modified 
the composition of the core by adding envelope material to the core region mimicking a diluted core. 
For extremely diluted cores with rock mass fraction below 20\% in the central region, 
they found a 50\% enhancement of the resulting atmospheric metallicity of Jupiter. 
Thus the relatively low outer envelope metallicities of our models suggest that diluted 
cores, possibly from core erosion, may be a better assumption for Jupiter than pure rock cores.

\section{Conclusions}\label{sec:concl}

We have aimed to push three-layer Jupiter models with our improved H EOS (H-REOS.2) 
in the direction of largest possible outer envelope metallicity $Z_1$. 
We find $Z_1 \lesssim 3.0\:\Zsol$, corresponding to an O:H enrichment factor $\lesssim 5$.
However, we prefer our models with $Z_1=2.0$--$2.5\:\Zsol$, corresponding to O:H=1--4~O:H$_{\rm solar}$, 
because they do not require to push all constraints to their $1\sigma$ uncertainty limits.
This increase in the envelope metal content compared to our earlier Jupiter models
based on the H-REOS.1 hydrogen EOS arises from about 2\% higher pressures in the 
0.2$-$1~{\gccm} and 5,000-10,000~K region of the H EOS, where the ab initio simulations 
are challenging since dissociation and ionization occurs. The resulting $Z_1$ values in 
our Jupiter models depend on the depth of the assumed transition to a helium-rich inner
envelope. The imposed constraint $Z_1> 2\:\Zsol$ requires a transition at $\Ptrans\geq 4\:$Mbar.
This threshold rises to $\Ptrans\geq 7\:$Mbar if fourth order terms in the rotational perturbation 
of Jupiter's gravity field and shape are included. Future investigations of the internal structure 
of Jupiter, and also of Saturn, should include a convergence check of the models with respect 
to the treatment of rotation.

Our two-layer model calculated with our EOS and our code differs from the \citet{Militzer+08} 
Jupiter model. We find a larger envelope metallicity ($4.5\:\Zsol$) that enables us to switch over 
to a three-layer model where the envelope is divided into a metal-poor outer and a metal-rich 
inner part to the expense of the core mass. 
In a low metallicity envelope (the M08 Jupiter model), this degree of freedom does not exist. 
The origin of the differences between these two-layer Jupiter models remains unexplained.  

For our three-layer models we calculate a cooling time of 4.4--5.0 Gyrs, an asymptotic frequency
spacing $\nu_0=156\pm 0.03\:\mu$Hz for global oscillation modes, and a normalized 
moment of inertia $\nI=0.276\pm 0.04\%$. These models are 
consistent with the available observational constraints. The Juno mission will be very
helpful for further constraining Jupiter's interior by measuring the deep atmospheric 
oxygen abundance, higher-order gravitational moments, and the moment of inertia.

\acknowledgments

The authors are grateful to the referee for the many comments and suggestions that significantly 
helped us to convert the initial manuscript into this paper. We acknowledge insightful discussions 
with J.~Fortney, T.~Guillot, and M.~French, and thank U.~Kramm for copy editing. 
This work was supported by the DFG RE 881/11-1, the DFG SFB 652,
the North-German-Supercomputing Alliance (HLRN) and the Computing Center of the University of Rostock.

\bibliographystyle{apj}
\bibliography{./ms-refs}

\appendix

\section{Method of calculating the entropy}\label{apx:entropy}

The functions $P(\rho,T)$ and $u(\rho,T)$ contain the full thermodynamic information.
We first re-construct the free energy $F$ and then use the Gibbs-Duhem relation
to convert to entropy.
It allows to derive the specific entropy $s(\rho,T)$ with an offset $s_0$
with respect to a reference state $(\rho_0,T_0$). The offset is unknown but constant for a 
given reference state. First, $F$ is derived with an unspecified offset $F_0=F(\rho_0,T_0)$ 
by integration of the total differential $d(F/T)$ along some path in $\rho-T$ space. $F$ is related 
to the entropy by definition, $F=U-TS$, where $U=M u$ and $S=M s$ are the extensive internal energy 
and the extensive entropy, respectively, and $M=V\rho$ is the mass contained in volume $V$. 
With the reference state $(T_0,V_0)$ and the unknown entropy offset $s_0$ we can write  
\begin{equation}\label{eq:STV}
	S(T,V) = \frac{U(T,V)}{T} -
\left(\frac{F(T,V)}{T}-\frac{F(T_0,V_0)}{T_0}\right) + s_0 M\:.
\end{equation}
The term in parenthesis in Eq.~(\ref{eq:STV}) is the solution to the line integral 
\begin{equation}
	\int_{T_0,V_0}^{T,V} d\textstyle\frac{F(T',V')}{T'}\quad,
\end{equation}
and independent of the chosen path of integration if the EOS data are thermodynamically
consistent. We choose lines of constant density and of constant temperature as our path of 
integration. While this is a choice of convenience it has the drawback of requiring a large scale 
EOS in $T-\rho$ space off the planetary adiabat. With $dU=-P\,dV + TdS$ we have
\begin{equation}
	\frac{d(U/T-S)}{dT}\:dT = - \frac{U}{T^2}\:dT\quad,\quad
	\frac{d(U/T-S)}{dV}\:dV = - \frac{P}{T}\:dV \quad.
\end{equation}
After switching back to $(\rho,T)$-space we can write
\begin{equation}\label{eq:dFTrho}
	\frac{1}{M}\int_{T_0,\rho_0}^{T,\rho}
d{\textstyle\frac{F(T',\rho')}{T'}} 
	= \int_{\rho_0}^{\rho}d\rho'\,\frac{1}{\rho'^2}\frac{P(T_0,\rho')}{T_0}
	- \int_{T_0}^{T}dT'\,\frac{u(T',\rho)}{T'^2}\:.
\end{equation}
After integrating the right hand side of Eq.~(\ref{eq:dFTrho}) we can numerically calculate the
entropy
\begin{equation}\label{eq:srhoT}
	s(\rho,T) = \frac{u(T,\rho)}{T} -
\frac{1}{M}\int_{T_0,\rho_0}^{T,\rho} d{\textstyle\frac{F(T',\rho')}{T'}} + s_0\:.
\end{equation}
If the underlying EOS is a mixture of different components, the entropy derived
from Eq.~(\ref{eq:srhoT}) implicitly includes the mixing terms, for instance the
ideal entropy of mixing if the EOS is an ideal mixture. The constant of integration
$s_0$ may depend on composition, but not on the initial condition of the
planetary adiabat. If a mixture is kept constant during a planet's evolution, 
the offsets $s_0$ of planetary adiabats through different outer boundaries
cancel each other, so that the real  entropy difference between two adiabats is
then known.

\section{Moment Of Inertia}\label{apx:moi}

The moment of inertia $I$ of a rigid body with spin-angular velocity $\omega$ and axis of rotation 
$\vec{\rm e}_\omega$ through the body's center of mass is
\begin{equation}\label{eq:I1}
I = \int_{\textsf V}d^3r \: \rho(\vec{r})\: (\vec{\rm e}_{\omega}\times \vec{r})^2\:,
\end{equation}
where the integral is taken over the volume $\textsf V$ of the body. If we choose the 
orientation of the body such that $\vec{\rm e}_{\omega}$ runs along the $z$-axis, then
$(\vec{\rm e}_{\omega}\times \vec{r})^2=r^2 \sin^2\vartheta$ in spherical coordinates.
Because for an isolated fluid planet the assumption of hydrostatic equilibrium implies symmetry 
around the axis of rotation, Eq.~(\ref{eq:I1}) becomes
\begin{equation}\label{eq:I2}
I = 2\pi \int_0^{\pi}\int_0^R d\vartheta dr \: \rho(r,\vartheta)\:r^4\:\sin^3\vartheta,
\end{equation} 
where $R$ is the planet's radius. If the planet rotates rapidly such as Jupiter, $R$
depends on the latitude $\vartheta$; in particular, the equatorial radius $R_{\rm eq}=R(\vartheta=\pi/2)$ 
is larger than the polar radius $R(\vartheta=0)$. Therefore, in order to calculate the two-dimensional 
integral the figure of the planet must be known. The figure is the shape at an iso-bar surface, 
equivalently to an equipotential surface. For the solar system giant planets, this surface is 
taken to be the 1~bar pressure level by convention. We have used the \emph{Theory of figures} \citep{ZT78}
to determine Jupiter's figure in terms of the figure functions $s_n(l)$ which give the shape of 
equipotential surfaces 
\begin{equation}\label{eq:r_l}
	r_l(l,\vartheta) = l\left(1+\sum_{n=0}^{\infty}s_{2n}(l)\:P_{2n}(\cos\vartheta)\right)\:.
\end{equation}
If the new radial coordinate $l$ is fixed, the total potential remains unchanged under 
variation of latitude $\vartheta$. The functions $P_{2n}(\cos\vartheta)$ 
are the Legendre polynomials. Only even expansion coefficients occur because of the assumed 
symmetry between northern and southern hemisphere. The sum in Eq.~(\ref{eq:r_l}) is truncated 
after the third order term $(n=3)$ because that is sufficient to calculate the gravitational 
moment $J_6$. We abbreviate the right hand side of the definition $r_l(l,\vartheta)$ by $r_l=l(1+\sum)$. 
By inserting Eq.~($\ref{eq:r_l}$) into Eq.~($\ref{eq:I2}$) and changing the independent 
variable from $r$ to $l$ using
\begin{equation}
	dr/dl = \left[\left(1+\sum s_{2n}\,P_{2n}\right) + l\left(\sum \frac{ds_{2n}}{dl} \,P_{2n}\right)\right]\,,
\end{equation}
and $l(P=1~{\rm bar})=R_{\rm m}$, the mean radius, we can finally compute $I$,
\begin{equation}
	I = 2\pi\int_0^{\pi}d\vartheta \,\sin^3\vartheta\: \int_0^{R_{\rm m}}dl
	\left[\left(1+{\textstyle \sum}\right) + l\left(\sum \frac{ds_{2n}}{dl} \,P_{2n}\right)\right]\:
	\rho(l) \: l^4(1+{\textstyle\sum})^4\:.
\end{equation}
In case of spherical symmetry, $\sum=0$ and $ds_{2n}/dl = 0$.

\clearpage

\clearpage

\clearpage

\clearpage

\clearpage

\clearpage

\clearpage

\clearpage
\input{tab1}

\end{document}

%% file: tab1.tex
\begin{deluxetable}{rrrrrrrr}
\tablewidth{0pt}
\tabletypesize{\small}
\tablecaption{\label{ascitab}Jupiter model J11-4a.}
\tablehead{
\colhead{$m$} & \colhead{$P$} & \colhead{$l$} & \colhead{$T$} & \colhead{$\rho$} 
& \colhead{$s_2$} & \colhead{$s_4$} & \colhead{$s_6$} \\
\colhead{($\ME$)} & \colhead{(GPa)} & \colhead{($\Rmean$)} & \colhead{(K)} & \colhead{(\gccm)}}
\startdata
317.833802 & 1.0000E-04 & 1.000000 &  170.0 & 1.6970E-04 &  -4.49641E-02 & 1.98582E-03 & -1.01995E-04\\
\vdots & \vdots & \vdots & \vdots & \vdots & \vdots & \vdots & \vdots\\
227.158225 & 3.9999E+02 & 0.735537 & 8995.9 & 1.3253E+00 &  -3.53291E-02 & 1.09438E-03 & -3.82050E-05\\
227.155808 & 4.0001E+02 & 0.735532 & 8996.0 & 1.4332E+00 &  -3.53290E-02 & 1.09437E-03 & -3.82046E-05\\
\vdots & \vdots & \vdots & \vdots & \vdots & \vdots & \vdots & \vdots\\
  7.567754 & 4.2729E+03 & 0.114019 & 20041.2 & 4.3140E+00 &  -1.00806E-02 & 8.05449E-05 & -6.72388E-07\\
  7.567666 & 4.2729E+03 & 0.114018 & 20041.2 & 1.8815E+01 &  -1.00805E-02 & 8.05447E-05 & -6.72385E-07\\
\enddata
\tablecomments{This table is published in its entirety in the electronic edition of the ApJ.
A portion is shown here for guidance regarding its form and content. The columns show the internal profile
of the Jupiter model J11-4a in terms of the mass coordinate $m$, the pressure $P$, the radial coordinate
$l$ scaled by Jupiter's calculated mean radius $\Rmean=10.95517\RE$, the temperature $T$, the density $\rho$,
and the figure functions $s_2$--$s_6$, see Appendix B.}
\end{deluxetable}